\documentstyle[times]{jaa}
\begin{document}
\title[Influence of Magnetic Field Decay on Electron Capture]{Influence of Magnetic Field Decay on Electron Capture in Magnetars}
\author[J. Zhang]%
       {Jie Zhang$^{1,2}$ \thanks{e-mail:zhangjie\_mail@cqu.edu.cn}\\
        1.Institute of Theoretical
Physics, China West Normal University, Nanchong  637009,
China\\2.Institute of Structure and Function, Chongqing University,
Chongqing 400044, China} \pubyear{2013} \volume{34}
%\pagerange{\pageref{217}--\pageref{219}}
%\setcounter{page}{217}
\date{Received 2013 Jan 28; accepted 2013 May 26}
\maketitle
\label{firstpage}
\begin{abstract}
The de-excited energy of electron capture (EC) induced by magnetic
field decay may be a new source for heating magnetar crust, so we
 do a quantitative calculation on the EC process near the outer crust and analyze their influence on persistent  X-ray radiation of magnetars, adopting the
experimental data or the results of theoretical model (including the
large-scale shell model and quasi-particle random phase
approximation).
\end{abstract}


\section{Introduction to the method}
Magnetars, neutron stars with ultrastrong magnetic field,  have been
 addressed by many researchers. In the recent years, the observations of magnetars suggest that
the luminosity of persistent X-ray radiated from magnetars is likely
the radiation of thermal
origin\nocite{5Ibr04}\nocite{6den08}\nocite{7Cam07}\nocite{8Tho02}\nocite{9Got06}(Ibrahim
et al. 2004; den Hartog et al. 2008; Camilo et al. 2007; Thompson et
al. 2007; G\"{o}tz et al. 2006). However, the considerable mechanism
of the X-ray source is not clear up to now. Cooper et al. proposed a
new heating mechanism in magnetar crusts\nocite{10Coo10} (Cooper and
Kaplan 2010). They argued that the magnetic pressure is comparable
to electron degeneracy pressure in magnetar crust, magnetic pressure
partially supports the crust against gravity. When the magnetic
pressure decreases and the crust shrinks, the density and electron
Fermi energy in crusts increase, then which induces exothermic
electron capture (EC) of nuclide (i.e., magnetic field-
decay-induced EC). In fact, the validity of this heating mechanism
strongly depends on the EC rate and the heat released in the EC
process.

Here we introduce our  quantitative results on this problem. We
employs a magnetar model with typical mass $M=1.4\, M_{ \odot}$ and
radius $R$=10 km\nocite{14Tho03} (Thompson 2003). The previous
researches show that, to avoid severe neutrino losses, the heat
source powering the thermal emission of the magnetar must be located
at or near the outer crust, i.e., within the magnetar's outermost
100 m. Hence we assume the outer crust's thickness is 0.1 km, which
is much less than its radius. We assume that the composition of the
crust is the final products of the rp-process\nocite{17Koi04} (Koike
et al. 2004). Of course, the products of rp-process may be quite
different due to the different accretion rate, ignition pressure of
nuclear burning and so on. We here choose Model 1a as an example, in
which $^{64}$Zn is the most abundant nuclides whose mass fraction is
34.7{\%}. Then it is easy to obtain the electron fraction $Y_{e
}\sim $ 0.48 according to the definition of electron fraction. Then
we estimate the average density $\bar {\rho }$ of crust by using the
hydrostatic equilibrium condition.

In our work, shell model are adopted to calculate the EC rates.
Strictly speaking, precise rate must consider all transitions from
the different initial state to different final state
\nocite{19Lan00}\nocite{20Pru03} (Langanke and Martinez-Pinedo 2000;
Pruet and Fuller 2003). However, it is unlikely to make an accurate
distribution for the all excited states of each nucleus because the
distribution of the high excited states is almost continuous
(particularly for the heavier nuclei). For the case of ground state
of parent nuclei, the nucleus spin and excited level distribution of
daughter nuclei can be found in the existing experimental data or
estimated by using the nuclear shell model. For the excited states
of parent nuclei, we adopted the results of large-scale shell-model
(LSSM, see Ref.\nocite{19Lan00}\nocite{21Lan99}(Langanke and
Martinez-Pinedo 1999; 2000)),  proton-neutron quasi-particle random
phase approximation (pn-QRPA) theory (see e. g., Nabi and Saijad
2008 \nocite{nab08}) or adopted the "Brink Hypothesis". LSSM's
calculations indicate that "Brink hypothesis" is valid for the bulk
of the GT strength \nocite{21Lan99}(Langanke and Martinez-Pinedo
1999). Since the surface temperature of magnetars are not high
enough (10$^{7}$-10$^{8}$K), most of parent nuclei are in the ground
state. Therefore ``brink hypothesis'' will not bring any substantial
deviation. We adopted all the level data whenever charge-exchange
experiment is available\nocite{25NN12} (NNDC 2012).

\section{Results}
In the initial stage of our model,   the electron chemical potential
in the crust is $3.68$ MeV. The EC threshold energy of the ground
state to ground state transition is defined as the mass of daughter
nucleus minus that of mother nucleus. So the negative threshold
energy indicates the EC reaction does not require additional
electronic energy; the positive threshold energy indicates that only
the electrons whose energy exceeds the corresponding threshold
energies can take place EC reaction effectively, that is, if
electron chemical potential is lower than the threshold energies,
only a small number of electrons in high-energy tail can take part
in the reaction. Certainly, their rates are very low. We find, for
most of nuclei, the  EC rate is dominant by the low energy
transition. For the most abundant nuclide, $^{64}$Zn, weighted mean
of de-excited energy $\bar {Q}(^{64}$Zn) = 0.42MeV. Since the life
of the magnetic field is much larger than that of $^{64}$Zn, most of
$^{64}$Zn will quickly decay into $^{64}$Cu. Because the electron
chemical potential is much higher than the EC threshold energy of
$^{64}$Cu, $^{64}$Cu will continue to capture electron quickly and
produce more stable $^{64}$Ni. Fortunately, the EC threshold energy
of $^{64}$Ni is 7.82MeV, which is much higher than the electron
chemical potential, so $^{64}$Ni is stable in this environment. A
similar analysis of other nuclides, we find $^{56}$Ni, $^{64}$Ga,
$^{60}$Ni and $^{55}$Co are also unstable, but their de-excited
energies are quit different. And we find the primary stable nuclides
are $^{56}$Fe, $^{64}$Ni, $^{60}$Fe, $^{55}$Cr and $^{12}$C, and the
electron fraction changes to 0.45.

As the magnetic field decreases, both the densities and electron
chemical potential will increase. However, we find the electron
electron chemical potential is still lower than the threshold
energies of most nuclei (except $^{56}$Fe, $^{56}$Fe$ \rightarrow
^{56}$Mn $\rightarrow ^{56}$Cr ($^{56}$Cr is stable)). This means
that, although the dynamics on the magnetic field decay will lead to
increased density and the electron chemical potential, the heat
released via EC induced by magnetic field decay is very limited for
the outermost crust, not as
large as the previous estimation. \\

%%%%%%%%%%%%%%%%%%%%%%%%%%%%%%%%%%%%%%%%%%%%%%%
\textbf{Acknowledgements}\\
%%%%%%%%%%%%%%%%%%%%%%%%%%%%%%%%%%%%%%%%%%%%%%%
 This work is supported by the National Natural
Science Foundation of China (grant 11273020) and the Science
Foundation of China West Normal University (grant 11B007).

%%%%%%%%%%%%%%%%%%%%%%%%%%%%%%%%%%%%%%%%%%%%%%%%%%%%%%%%%%%%%%%%%%

\end{document}